\documentclass[runningheads]{llncs}
\usepackage{graphicx}
\usepackage{url}
\usepackage{hyperref} 
\usepackage{threeparttable}
\usepackage{framed,multirow}
\usepackage{multirow}
\usepackage{gensymb}
\usepackage{hhline}
\usepackage{rotating,graphicx}
\usepackage{ulem}
\usepackage{babel}
\usepackage{amsmath}
\usepackage{mathtools}
\usepackage{graphicx}
\usepackage{fancyhdr}
\usepackage{hyperref}
\usepackage{tabularx, booktabs}
\usepackage[ruled,vlined]{algorithm2e}
\usepackage{algorithmic}
\usepackage{caption}
\usepackage{subcaption}
\begin{document}
\sloppy
\title{Learning to Rearrange Voxels in Binary Segmentation Masks for Smooth Manifold Triangulation}
%
%
\author{Jianning Li\inst{1,2,3,4}\and
Antonio Pepe\inst{1,2}\and
Christina Gsaxner\inst{1,2} \and
Yuan Jin\inst{1,2,5}  \and
Jan Egger \inst{1,2,4}
} 

\authorrunning{Learning Smooth Manifold Triangulation}
%
\institute{Institute of Computer Graphics and Vision, Graz University of Technology, Inffeldgasse 16, 8010 Graz, Austria \and
Computer Algorithms for Medicine Laboratory (Caf\'e-Lab), 8010 Graz, Austria. \and
Research Unit Experimental Neurotraumatology, Department of Neurosurgery, Medical University Graz, Auenbruggerplatz 2$^{2}$, 8036 Graz, Austria. \and
Institute for AI in Medicine (IKIM), University Hospital Essen, Girardetstraße 2, 45131 Essen, Germany \and 
Research Center for Connected Healthcare Big Data, ZhejiangLab, Hangzhou, Zhejiang, 311121 China\\
\email{\{jianning.li,egger\}@icg.tugraz.at}}

\maketitle              
\begin{abstract}
Medical images, especially volumetric images, are of high resolution and often exceed the capacity of standard desktop GPUs. As a result, most deep learning-based medical image analysis tasks require the input images to be downsampled, often substantially, before these can be fed to a neural network. However, downsampling can lead to a loss of image quality, which is undesirable especially in reconstruction tasks, where the fine geometric details need to be preserved. In this paper, we propose that high-resolution images can be reconstructed in a coarse-to-fine fashion, where a deep learning algorithm is only responsible for generating a coarse representation of the image, which consumes moderate GPU memory. For producing the high-resolution outcome, we propose two novel methods: learned voxel rearrangement of the coarse output and hierarchical image synthesis. Compared to the coarse output, the high-resolution counterpart allows for smooth surface triangulation, which can be 3D-printed in the highest possible quality. Experiments of this paper are carried out on the dataset of AutoImplant 2021 (\url{https://autoimplant2021.grand-challenge.org/}), a MICCAI challenge on cranial implant design. The dataset contains high-resolution skulls that can be viewed as 2D manifolds embedded in a 3D space. Codes associated with this study can be accessed at \url{https://github.com/Jianningli/voxel_rearrangement}.    

\end{abstract}
\section{Introduction}
\subsection{Background}
For many medical image analysis tasks, the high-resolution images, especially 3D images, have to be downsampled before they can be fed into deep neural networks, due to limited GPU capacity. However, downsampling can often lead to severe degradation of image quality and loss of subtle structures, which is unacceptable, especially in precision-demanding reconstruction tasks.

\subsection{Related Work}
In tackling the problem of high memory consumption of large 3D medical images, the medical image analysis community has come up with several techniques, which we roughly grouped into three categories based on which part of the image analysis pipeline these techniques have targeted: the medical image itself, the network architecture or the data structure. 

\subsubsection{Medical Images}
Dividing the high-resolution medical images into smaller patches is the most intuitive and prevalently adopted method to fit the image to the available GPU memory. If necessary, the strategies of cropping patches from an image have to be adapted to the characteristics of the data. For example, Li, J. et al. \cite{li2021automatic} proposed to train a deep neural network using successive non-overlapping patches and overlapping patches, so that the network can learn the global shape distribution of the high-resolution, spatially sparse skull data effectively. Akil, M. et al. \cite{akil2020fully} extracted overlapping patches from MRI images for brain tumor segmentation, considering that using overlapping patches cropped from the images helps the deep neural network to learn the spatial relationship among patches. 

Other researchers have investigated detail-preserving image downsampling techniques \cite{diaz2017downsampling}, which allow for the reduction of image size without substantially degrading the image quality. 

\subsubsection{Network Architecture}
Two popular network structures dealing with high-resolution medical images include the \textit{coarse-to-fine} framework and the cascaded neural network. As the name suggests, under the \textit{coarse-to-fine} framework, a network first produces a coarse image output, which is further refined to approximate the ground truth. In \cite{li2020baseline}, the authors first trained an autoencoer on downsampled skull images. The corresponding coarse output is used to extract the region of interest (ROI) of the original high-resolution defective skull. The size of the ROI, compared to the whole image, is reduced substantially, while no details essential to the specific task have been discarded. The ROI is then used to train another autoencoder to produce the final fine output. In \cite{han2017high}, in order to perform shape completion on high-resolution ($256^3$) grids, the authors first trained a neural network on the coarse counterpart of the object ($32^3$), which, even if is of low resolution, contains the global shape information of an object. The corresponding learnt feature maps are used to guide the training of another neural network, which works on patches cropped along the missing region of the original high-resolution object ($256^3$). The work of Dai, A. et al \cite{dai2017shape} is the closest to our study, in which the coarse output of the first network is hierarchically synthesized to higher resolution, based on an image template.   

Kodym, O. et al. \cite{kodym2020skull,kodym2020cranial} used cascaded convolutional neural networks (CNN) for skull shape completion, where the first network (3D U-net) takes as input downsampled skull volumes ($64^3$) and produces a coarse output. The second network takes as input the preceding output (upsampled to $128^3$) as well as an equally sized patch cropped from the original high-resolution skull and produces the final high-resolution output.   

\subsubsection{Data Structure}
Some medical images (e.g., the skull) have unique characteristics such that a memory-efficient data structure can be tailor-made for them. Graham, B. et al. have devised \textit{Submanifold Sparse CNN} \cite{graham2017submanifold,graham20183d}, a set of convolutional operations tailored for processing spatially sparse data in the form of submanifolds. Different from conventional convolutional filters that slide over the entire image space, including the object of interest and the background, sparse convolutions run only on the object. This is efficient, memory-saving and requires substantially fewer floating point operations (FLOPs), especially when the object occupies only a small percentage of the entire volume (like the skull, which is essentially a two-dimensional surface in a 3D volumetric space, i.e., a manifold). Similarly, Riegler, G. et al. \cite{riegler2017octnet} devised \textit{OctNet}, which can learn shape representations from high-resolution data stored in an Octree data structure.

\section{Dataset}
The datasets for \textit{Task 3} of the AutoImplant 2021 challenge (\url{https://autoimplant2021.grand-challenge.org/}) were used. They account for 100 triplets of complete skulls, artificial defective skulls and the corresponding implants for training and 110 triplets for testing\footnote{The test set is further split into $D_{100}$, which contains 100 defective skulls with similar defect shapes to those of the training set and $D_{10}$, which includes 10 defective skulls with varied defect patterns.}. According to \cite{li2020dataset}, the skulls are binary segmentation masks with a very low voxel occupancy rate (VOR, usually less than 10 percent), indicating that in the large volumetric space, $512\times512\times Z$, only a fraction of the voxels contain valid geometric information of the skulls. Existent deep learning-based methods usually take as input the entire volume, which consumes an excessive amount of memory and computation power \cite{li2021autoimplant}. Besides sparsity, another characteristic of the skull voxels is that they are distributed approximately spherically (see Fig. 8 in \cite{li2020baseline} as an example) in the volumetric space, where the non-zero voxels constitute the surface of the skull, making the skull resemble a manifold, topologically.

\section{Method}
An extensive review of the solutions presented at the AutoImplant challenge~\cite{li2021autoimplant} in 2020 has identified two major difficulties of the deep learning-based cranial implant design task: \textbf{1)} The skull volumes are of high resolution and a large amount of memory and computation power is needed to process them. With limited computation resources (especially within clinical settings~\cite{chang2021three}), the skulls have to be downsampled and, subsequently, the output of networks is coarse. \textbf{2)} In high resolution, deep learning models tend not to generalize well across varied defect patterns. 

\begin{figure*}[ht]
     \centering
        \includegraphics[width=\textwidth]{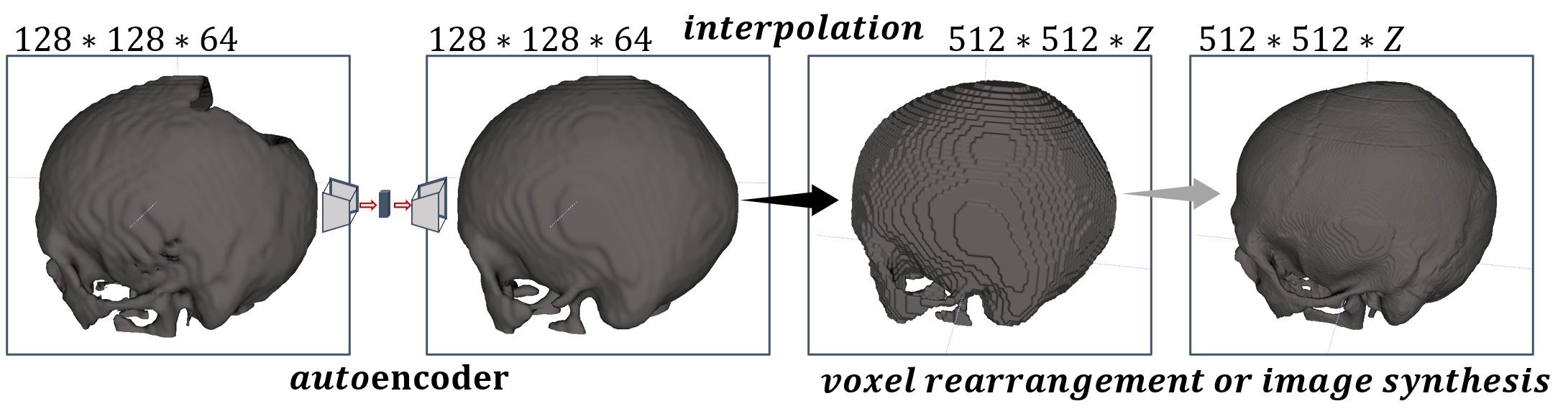}
     \caption{Pipeline of the proposed coarse-to-fine framework for the high-resolution skull shape completion task.}
     \label{fig:pipeline}
\end{figure*}

According to one of our earlier works~\cite{li2020baseline} (Appendix A and B), a simple autoencoder shows good generalization performance for shape completion of varied defect patterns when trained on low-resolution (downsampled) skulls, without an augmented training set (only the original 100 skull pairs in the training set provided by the AutoImplant challenge were used). Based on these findings, we propose a coarse-to-fine framework, as illustrated in \autoref{fig:pipeline}, where the skull shape completion is carried out on downsampled skulls ($128\times128\times64$) using an autoencoder network. We use the same autoencoder as the skull shape completion network used in~\cite{li2020baseline}, i.e., $N_1$. The low-resolution output is then upsampled to its original resolution, $512\times512\times Z$, using spline interpolation. This results in a coarse upsampled skull, as shown in \autoref{fig:pipeline}. As a final step, the geometric details and smoothness of the skull surface is restored through either voxel rearrangement or image synthesis. A detailed description of voxel rearrangement and image synthesis follows.

\subsection{Learning Voxel Rearrangement}
\begin{figure*}[ht]
     \centering
        \includegraphics[width=\textwidth]{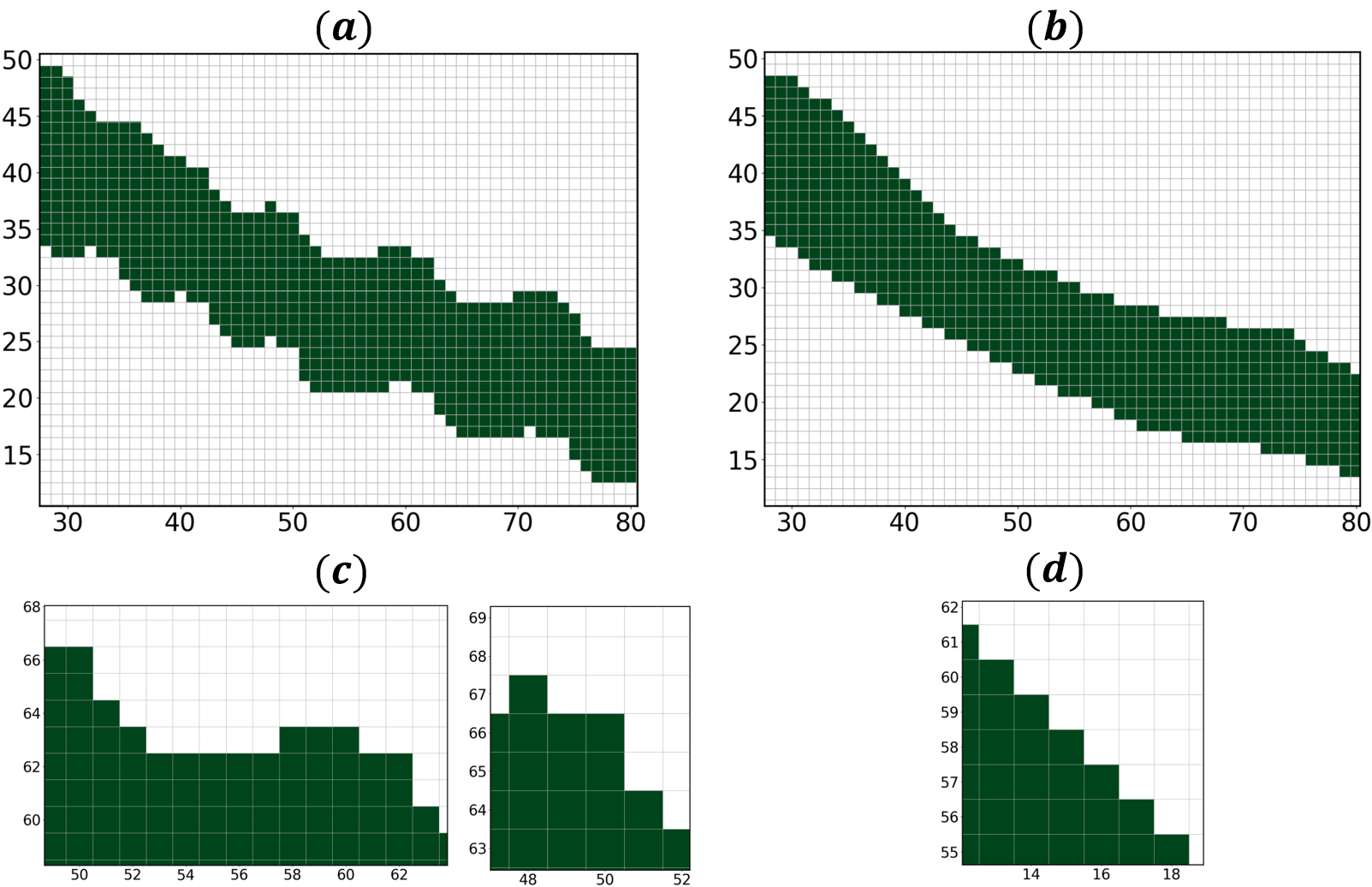}
     \caption{Illustration of voxel arrangement on the surface of coarse (a) and ground truth (b) skulls. (c) and (d) show the dominant voxel arrangement patterns on the two types of skulls.}
     \label{fig:voxel_arrange}
\end{figure*}

In this study, it is assumed that, through rearrangement of the empty and non-empty voxels, the surface of a coarse skull can be smoothed and the geometric details can be restored. It should be noted that the surface smoothness obtained via voxel rearrangement substantially differs from that obtained using smoothing filters such as median or Gaussian filters, which are not detail-preserving filters.     
\autoref{fig:voxel_arrange} shows an illustration of the (inner and outer) surfaces of a coarse skull (a) and the ground truth skull (b), from which we can see the difference of the voxel arrangements between the two. For ease of illustration, we used a 2D grid to represent the volumetric data, where the filled grid cells stand for occupied voxels (valued 1) and the blank cells stand for background (valued 0). \autoref{fig:voxel_arrange} (c) and (d) show, respectively, the dominant voxel arrangement patterns on the surface of the coarse skull and the ground truth skull. We can see that, for the ground truth skull in (d), the neighboring occupied voxels tend to be arranged in descending terraces and the step size is only one voxel. Besides, when seeing the occupied voxels in the grid as a curve, the sign of the first derivative is consistent.
In comparison, the step size for the occupied voxels on the coarse skull is usually larger, e.g., two voxels, and the derivative sign can change locally. 

\begin{figure*}[ht]
     \centering
        \includegraphics[width=\textwidth]{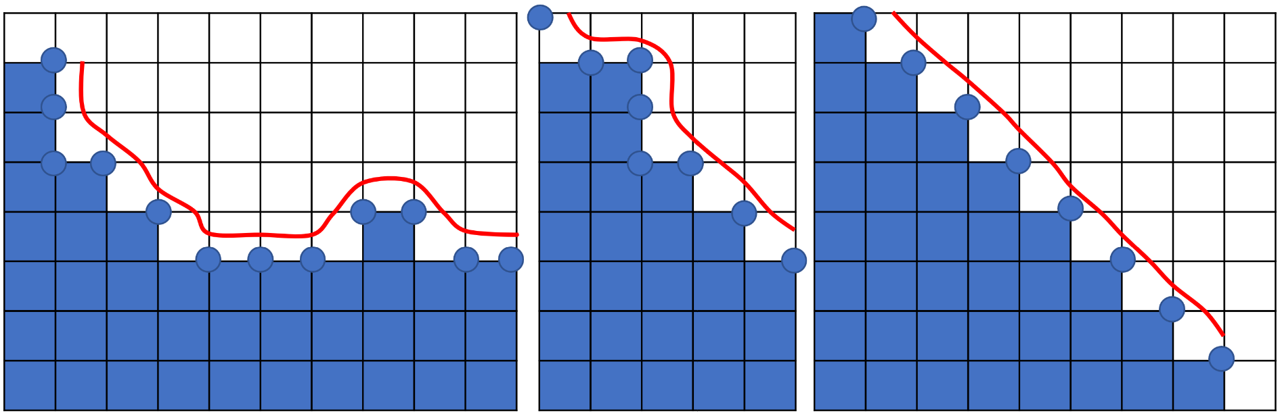}
     \caption{Polygon mesh (red) extracted from the isosurface of the binary skull volume. Left and middle: the coarse skull. Right: the ground truth.}
     \label{fig:marching_cube}
\end{figure*}

Differe voxel arrangement patterns can result in different triangulation results on the binary masks, as illustrated in \autoref{fig:marching_cube}, where the occupied voxels are depicted using blue dots on the vertices of the 2D grid. Taking the marching cubes algorithm as example, which extracts a polygon mesh from the isosurface (the skull surface formed by all the occupied voxels), whether or not an edge of the mesh passes through a cube on the grid is determined by whether there is one edge of the cube that contains two opposite (zero and one) vertices. As previously mentioned, if one vertex of the cube on the grid is one, the vertex belongs to the object and vice versa. Based on this rule, (a segment of) the polygonal surfaces for the three voxel arrangement patterns can be obtained and are shown in red in \autoref{fig:marching_cube}.

We can see that, with a large step size (e.g., of two voxels) and an inconsistent derivative sign of the isosurface, the resultant polygon edges are locally not smooth (e.g., they have bumps or \textit{'sharp'} turns). In contrast, the polygon from the ground truth is smooth (\autoref{fig:marching_cube}, right). It is easy to extend the above concept to 3D, where a skull volume is represented by a 3D grid containing $512\times512\times Z$ cubes.

In our study, we propose to learn a transformation $\mathbf{F}$ from a coarse voxel arrangement (\autoref{fig:voxel_arrange} (a)) to a smooth arrangement (\autoref{fig:voxel_arrange} (b)), using an autoencoder network. The same autoencoder used for low-resolution skull shape completion is used for the voxel rearrangement, where the input is the the upsampled coarse skull and the ground truth is the original complete skull ($512\times512\times Z$). For each training epoch, $128\times128\times 64$ patches are randomly cropped from the two types of skulls. During inference, as the number of slices $Z$ of each skull differs, we first sequentially feed the upper $\frac{512}{128} \times \frac{512}{128} \times (Z // 64)$ patches ($//$ operator takes only the integer after a division) of the image into the trained network. Then, the patches are cropped from the lower 64 slices. The final result is obtained by stitching the output patches in the same sequence as they are fed into the network together. Doing so allows us to handle varied-sized skull volumes. 

As both the input and ground truth are binary, learning voxel rearrangement (rearranging the positions of empty and non-empty voxels) can also be viewed as learning a conversion between 0 and 1:

\begin{equation}
\begin{cases}
\mathbf{F}(0)=1 & \text{ if } \chi_1\\ 
\mathbf{F}(0)=0 & \text{ if } \chi_2 \\ 
\mathbf{F}(1)=1 & \text{ if } \chi_3\\ 
\mathbf{F}(1)=0 & \text{ if } \chi_4
\end{cases}
\label{eq: voxel_rearrangement}
\end{equation}

The network learns a set of rules ${\chi_1, \chi_2,\chi_3,\chi_4}$ in order to determine which conversion described in Equation \ref{eq: voxel_rearrangement} will be executed during the inference for each voxel position.

\begin{figure*}
     \centering
        \includegraphics[width=\textwidth]{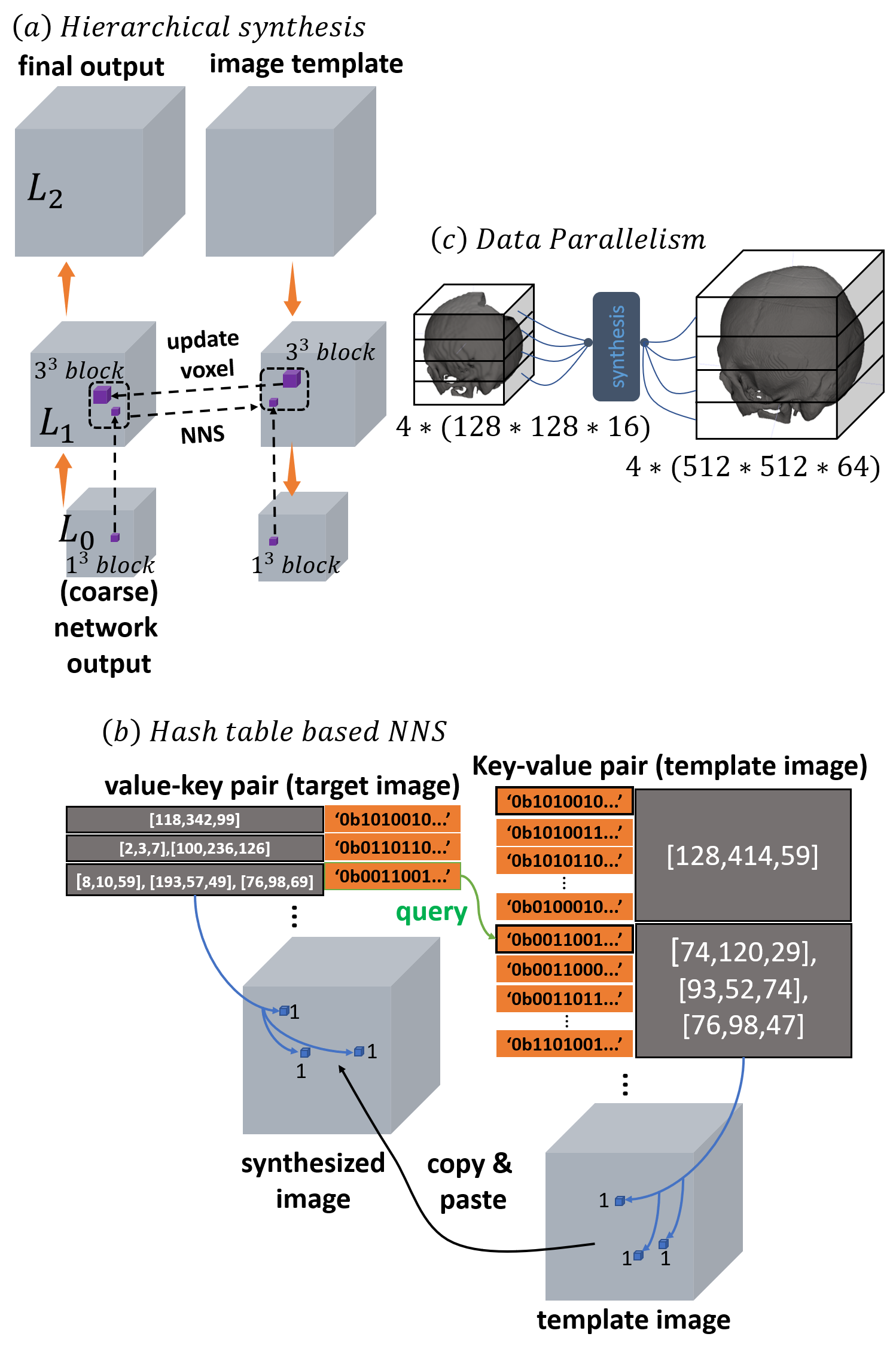}
     \caption{Illustration of voxel updating based on a hierarchical image synthesis pipeline, which consists of three main components: (a) the creation of a hierarchical pyramid, (b) the creation of hash tables and binary encoding, and (c) data parallelism.}
     \label{fig:image_synthesis}
\end{figure*}

\subsection{Hierarchical Image Synthesis}
Besides using a neural network to learn a smooth voxel arrangement as described in Section 3.1, the voxels in a coarse skull can also be updated according to a pre-selected smooth skull template. In this section, we introduce a scheme where the coarse skull output (from the skull shape completion autoencoder trained on downsampled skull data) can be hierarchically synthesized to high resolution, as illustrated in \autoref{fig:image_synthesis} (a). For a fast and memory-efficient synthesis, a tailored hash table-based approximate nearest neighbor search (NNS) strategy and binary encoding is used (\autoref{fig:image_synthesis} (b)). To further speed up the process, we divide each skull into four sub-volumes and the synthesis of each sub-volume is parallelized by making use of the multicore processing potential of the CPU (\autoref{fig:image_synthesis} (c)).  

\subsubsection{Hierarchical Image Synthesis}
As shown in \autoref{fig:image_synthesis} (a), Gaussian image pyramids are created for the coarse network output\footnote{The coarse output (size: $128 \times 128 \times 64$) by the shape completion autoencoder network shown in \autoref{fig:pipeline}.}, and a randomly selected complete skull from the training set, which will serve as a smooth template. Starting from the bottom pyramid level $L_0$, each voxel in the first ($L_1$, $256 \times 256 \times 128$) and second level ($L_2$, $512 \times 512 \times 256$) of the coarse output image pyramid is replaced by its most similar voxel in the corresponding level of the smooth template image pyramid. The similarity between two voxels, $V1$ and $V2$, is measured by the Hamming distance between the neighbors of the voxels:

\begin{equation}
d_{HM}(V1,V2)= \sum(V1_n \bigoplus V2_n)
\label{eq: hamming_dist}
\end{equation}

$V1_n$ and $V2_n$ are the $3^3$ or $5^3$ neighbors of the two voxels. Here, a $3^3$ neighborhood is used. Considering that the images are binary, we used a bit representation of the voxels and their neighbors. For example, the $3^3$ neighborhood of a voxel is stored as a 27-bit long string such as \textit{'ob111001011100...'}. This reduces the memory consumption substantially compared to using the original data type, where each voxel occupies 64 (\textit{int64}) or 32 (\textit{int32}) bits of memory. $\bigoplus$ stands for the bit-wise XOR of the two bit strings. $\sum$ counts the number of non-empty bits.

\subsubsection{Sparsity and Approximate NNS}
To upsample a coarse skull to a higher resolution, all the voxels in the coarse pyramid have to be updated based on their similarity to the voxels in the template pyramid, which can be formulated as a nearest neighbor search problem. However, using a linear search strategy is impractical in the situation, considering that the number of voxels in a skull volume is usually large, e.g., there are over 60 million voxels in a $512 \times 512 \times 256$ volume. To reduce the required number of searches, we take the sparsity of the skull volume into consideration. For both, the coarse pyramid and the template pyramid, only the voxels whose neighbors are non-empty are involved in the NNS\footnote{For example, in a typical $512 \times 512 \times 256$ volume containing the skull, there may only be around three million of such voxels, which is, however, still impractical for a linear search, as over three million $\times$ three million comparisons are needed to update all voxels.}. In other words, we only update the voxels in and around the skull surface.

Similar to the implementation of sparse convolutional operations described in \cite{graham2017submanifold,graham20183d}, we use a hash table to store the bit strings (i.e., the voxels and neighbors) and the corresponding coordinates $(x,y,z)$ for both the coarse pyramid and the template pyramid, as shown in \autoref{fig:image_synthesis} (b). It should be noted that one bit string (the \textit{key} in the hash table) could correspond to multiple coordinates, when several voxels have the same neighbors. In such cases, only a one-time search is required for these voxels, which further reduces the overall number of searches needed.

No matter how large a hash table is, the time complexity of retrieving an entry from a hash table is always
$O(1)$, which is a highly desirable property for our task, provided that an entry from the coarse pyramid exists also in the template pyramid (i.e., the Hamming distance is zero). In order to increase the likelihood that a bit string exists in both hash tables, we pre-compute all the bit strings that have an Hamming distance (calculated according to Equation \ref{eq: hamming_dist}) below three from each actual \textit{key} in the template pyramid. These string neighbors correspond to the same coordinate as the actual \textit{key} that exists in the template pyramid. Experimentally, a set of such neighbors\footnote{Not to be confused with the $3^3$ and $5^3$ voxel neighbors in the image pyramid.} together with the actual keys would guarantee that one entry from the coarse pyramid exists also in the template pyramid with a probability of over 80\%. For the remaining voxels, zero and one can be randomly assigned. 

\begin{figure}[ht]
  \centering
  \begin{minipage}{.7\linewidth}
   \begin{algorithm}[H]
   \caption{Retrieving the coordinates corresponding to an entry (bit string) from a hash table}
   \textbf{Input:} a \textit{key} $K_c$ from the coarse pyramid   \;
    \textbf{Output:} coordinate(s) $(x,y,z)$ from the template pyramid  \;
  \uIf{$K_c$ in $S_{ta}$}{
    coordinates=$S_{ta}$.get\_value$(K_c)$ \;
  }
  \uElseIf{$K_c$ in $S_{tn}$}{
    coordinates=$S_{tn}$.get\_value($K_c$) \;
  }
  \Else{
    assign 0 or 1 to the voxels \;
  }
  \label{alg:algo_1}
\end{algorithm}
  \end{minipage}
\end{figure}

Define the sets of actual keys and their neighbors from the template image as $S_{ta}=\left \{K1_{actual}, K2_{actual}, K3_{actual} \cdot\cdot\cdot\cdot \right \}$ and $S_{tn}=\left \{K_{n1}, K_{n2}, K_{n3} \cdot\cdot\cdot\cdot \right \}$, given a \textit{key} $K_c$ from the coarse pyramid, the coordinates of the voxels in the template pyramid to be used to replace the voxels in the coarse pyramid can be obtained according to \autoref{alg:algo_1}. 
The process is extremely fast in runtime, even for large hash tables, as the time complexity of the two main operations (i.e., get\_value) is constant $O(1)$. 

\subsubsection{Data Parallelism}

By making use of the multicore processing potential of the CPU, each skull volume can be divided into four equal patches and the synthesis of the patches can be parallelized, as illustrated in \autoref{fig:image_synthesis} (c). Note that if data parallelism is used, the hash tables should also be created for patches, instead of on the whole skull volume. The final output is the combination of the four output synthesized patches.

\section{Experiment and Results}
As described in Section 3, an autoencoder is first trained on downsampled skulls ($128 \times 128 \times 64$) for skull shape completion. We evaluate three approaches for upsampling the coarse output to the original resolution ($512 \times 512 \times Z$): spline interpolation, voxel rearrangement (proposed) and hierarchical image synthesis (proposed).

For the interpolation-based upsampling, the completed, coarse skulls produced by the trained shape completion network are upsampled to their original size of $512 \times 512 \times Z$ using spline interpolation.

For the proposed voxel rearrangement-based method, the trained autoencoder runs first on the defective skulls in both the training and test set to produce the corresponding completed and coarse skulls. The predicted coarse skulls are then upsampled to $512 \times 512 \times Z$ using spline interpolation. To learn a smooth voxel arrangement, another autoencoder (the same as the autoencoder used for skull shape completion in the first step) takes as input a random $128 \times 128 \times 64$ patch cropped from the upsampled coarse skull. The ground truth is the corresponding patch cropped from the complete skull in the training set. After training, the autoencoder runs on the upsampled coarse skulls in the test set to produce the final completed and voxel-rearranged high-resolution skulls. 

For comparison with the above approaches, an autoencoder network is also trained using randomly cropped patches for patch-wise shape completion. From the training set, the autoencoder takes a randomly cropped patch from an original, high-resolution, defective skull as input and produces a complete skull patch as ouptut. For a fair comparison, the training and inference strategy used in the patch-wise shape completion is the same as that of the voxel rearrangement-based method (as described in Section 3.1), in order to handle differently sized skull data. The autoencoder network used here is also the same as that used in the interpolation and voxel rearrangement-based method. 

\subsection{Interpolation, Patch-wise skull shape completion and Voxel rearrangement}

For the three approaches (interpolation, patch-wise skull shape completion and voxel rearrangement), the implants are obtained by subtracting the defective skulls in the test set from the final complete skulls generated by the algorithms. Experiments were carried out on the training and test set ($D_{100}$ and $D_{10}$) of \textit{Task 3} of the AutoImplant 2021 Challenge.

\begin{table*}[ht]
\centering
\caption{Mean values of the Dice Similarity Coefficient (DSC) and the Hausdorff Distance (HD, measured in mm) for the skulls and implants on the test set of \textit{Task 3}.}
\begin{tabular}[t]{ccccccccccccc} 
\toprule
 \multirow{2}{*}{Methods}  & \multicolumn{3}{c}{\textbf{skull($D_{100}$)}} & \multicolumn{3}{c}{\textbf{skull ($D_{10}$)}} & \multicolumn{3}{c}{\textbf{implant ($D_{100}$)}} & \multicolumn{3}{c}{\textbf{implant ($D_{10})$}}\\ 
 \cmidrule{2-3}  \cmidrule{5-6} \cmidrule{8-9} \cmidrule{11-12}

                        & $DSC$            & $HD$            &  & $DSC$            & $HD$  &       & $DSC$             & $HD$ &    & $DSC$         &$HD$ \\ 
\hline
Interpolation              &0.7547          & 24.4227   &           &0.7546        & 23.5864         &   &0.8151 &32.5061 & &0.7135 &42.3458   \\
Voxel Rearrangement              &0.7529         &37.1932             &        & 0.7574       & 24.9146  & &0.8135 &30.8189 & &0.7563 & 28.0752    \\
Patch              &0.8587          &16.8571               &       &0.8493          &27.0759    & &0.6178 &27.0386 &  &$-$ & $-$ \\

\bottomrule
\end{tabular}
\label{table:table1}
\end{table*}

\begin{figure}
     \centering
     \begin{subfigure}[b]{\textwidth}
         \centering
         \includegraphics[width=\textwidth]{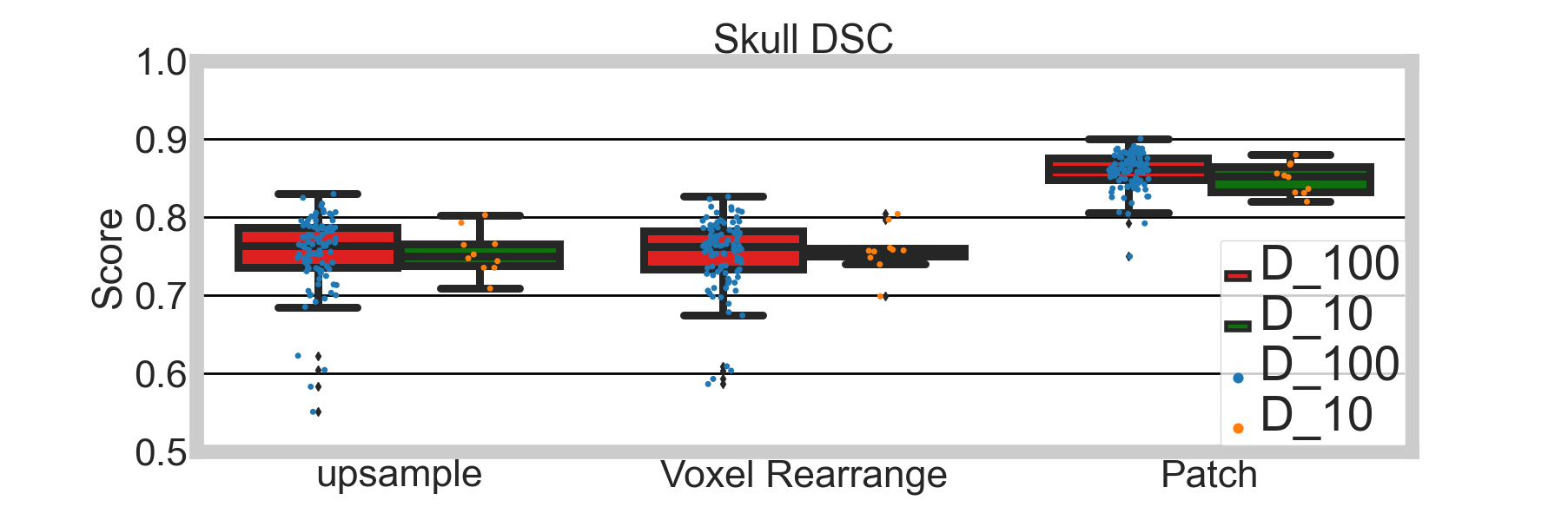}
         \caption{DSC of the skull}
         \label{fig:dsc_skull}
     \end{subfigure}
     \hfill
     \begin{subfigure}[b]{\textwidth}
         \centering
         \includegraphics[width=\textwidth]{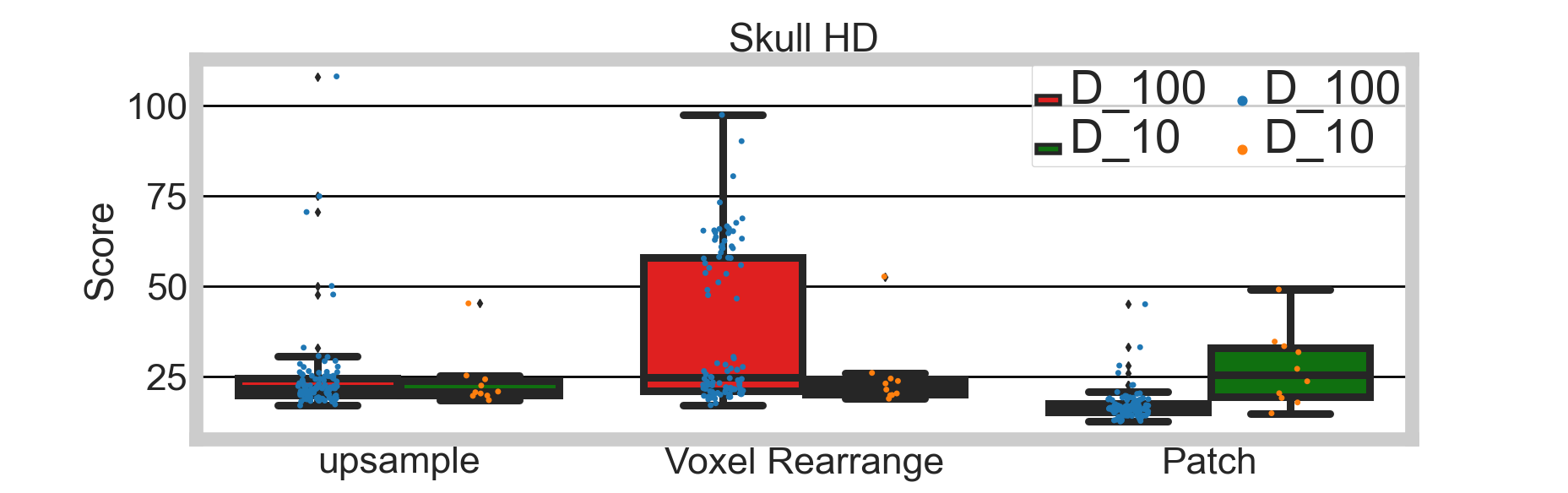}
         \caption{HD of the skull}
         \label{fig:hd_skull}
     \end{subfigure}
     \hfill
     \begin{subfigure}[b]{\textwidth}
         \centering
         \includegraphics[width=\textwidth]{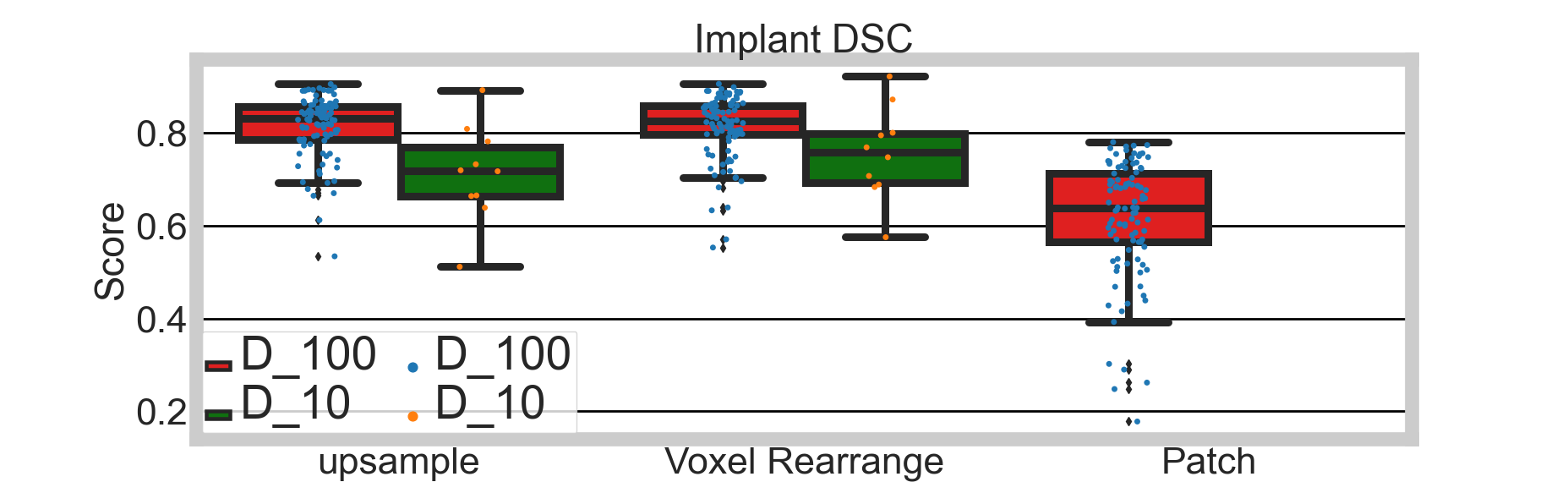}
         \caption{DSC of the implant}
         \label{fig:dsc_implant}
     \end{subfigure}
     
      \begin{subfigure}[b]{\textwidth}
         \centering
         \includegraphics[width=\textwidth]{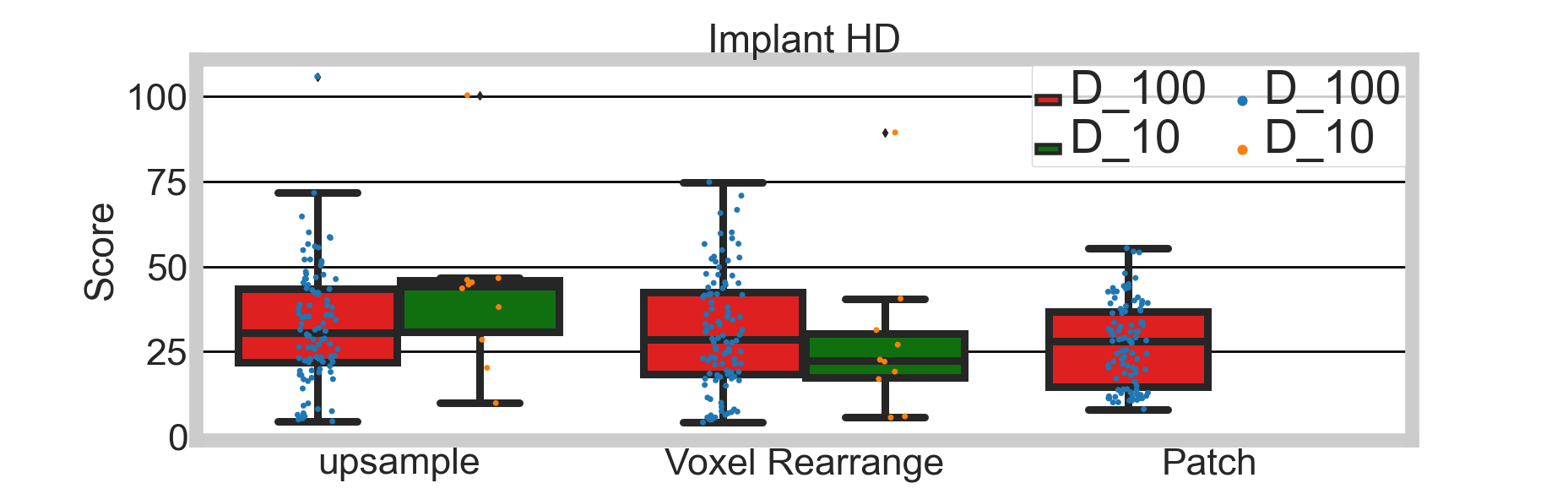}
         \caption{HD of the implant}
         \label{fig:hd_implant}
     \end{subfigure}    
     
        \caption{Quantitative comparison (DSC, HD) of the three methods (upsample via spine interpolation, voxel rearrangement and patch-wise skull shape completion) on the test sets ($D_{100}$, $D_{10}$) of Task 3 of the AutoImplant challenge.}
        \label{fig:dsc_hd}
\end{figure}

Table \ref{table:table1} shows the Dice similarity Coefficient (DSC) and the Hausdorff Distance (HD) of the completed skulls and implants for the three approaches: spline interpolation, voxel rearrangement and patch-wise skull shape completion on the test set ($D_{100}$ and $D_{10}$). \autoref{fig:dsc_hd} (a - d) shows the corresponding boxplots.   

\begin{figure*}[ht]
     \centering
        \includegraphics[width=\textwidth]{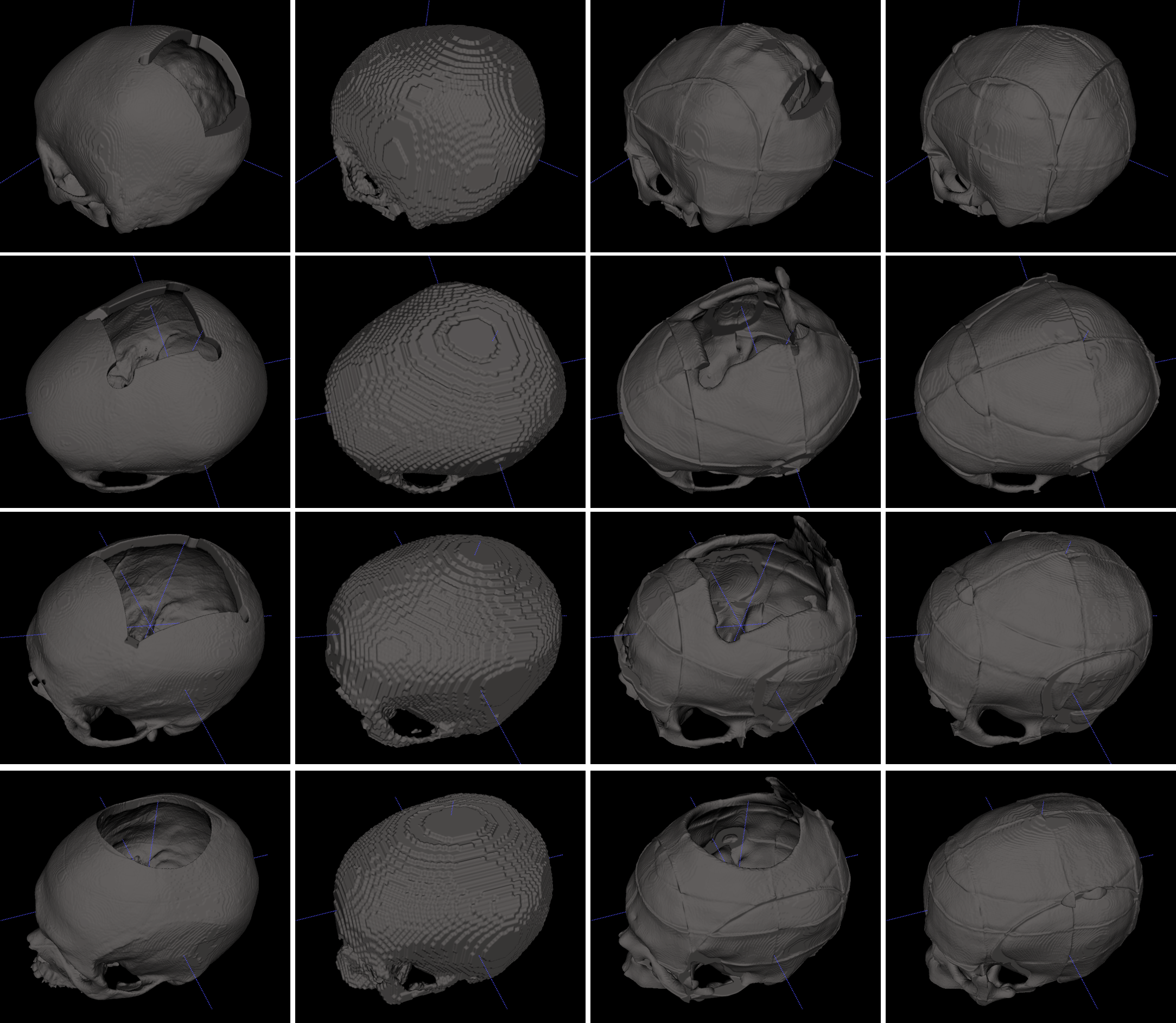}
     \caption{From first to fourth column: the input defective skull, the completed skull from spline interpolation, patch-wise skull shape completion and voxel rearrangement (proposed) .}
     \label{fig:result_skull}
\end{figure*}

\autoref{fig:result_skull} shows the skull shape completion results on $D_{100}$ (first row) and $D_{10}$ (second to fourth row) for the three approaches. We can see that even if quantitatively, the patch-wise skull shape completion method has the best scores regarding DSC and HD for the skull, the qualitative inspection of the completed skulls reveals that the patch-based completion method failed on $D_{10}$ and cannot completely restore the missing parts of the skulls on $D_{100}$. On the contrary, the autoencoder trained on downsampled skulls shows good generalization performance on both $D_{100}$ and $D_{10}$. For the skulls, despite that the DSC for the interpolation-based upsampling is quite close to that of the voxel rearrangement-based upsampling and that the HD of the interpolation-based upsampling is even smaller than that of the voxel rearrangement-based upsampling, \autoref{fig:result_skull} clearly shows the advantages of the voxel rearrangement-based upsampling in terms of the reconstruction quality (mainly the skull surface) between the two approaches. It should be noted that, due to the patch-wise training and inference scheme used for voxel rearrangement, we can see \textit{stitching} lines on the contacting borders between neighboring patches (\autoref{fig:result_skull}, last column), which is undesirable.

\begin{figure*}[ht]
     \centering
        \includegraphics[width=\textwidth]{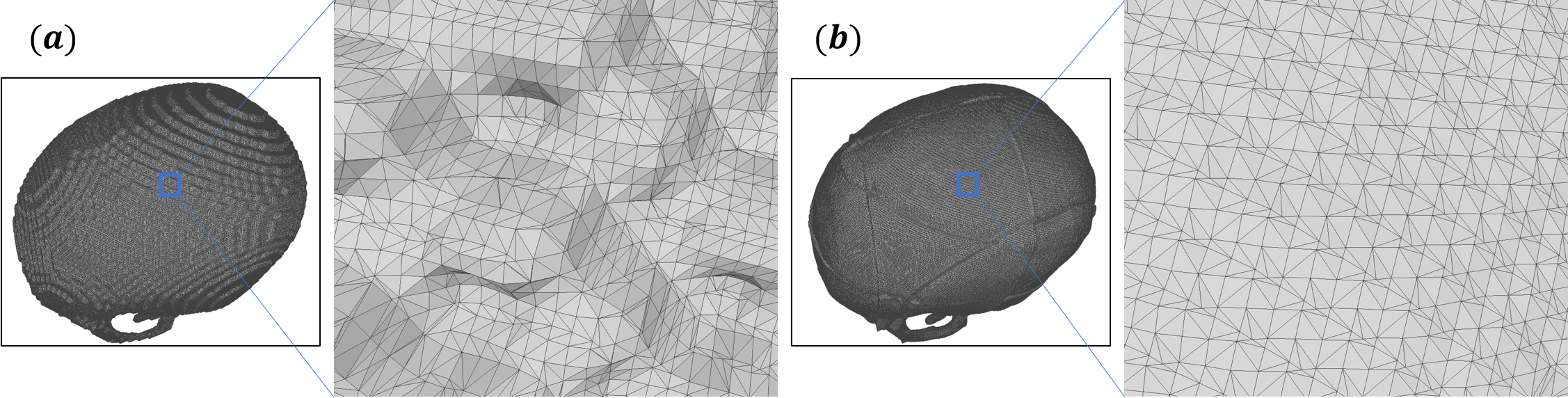}
     \caption{Triangulation of the skull voxel grids produced by the interpolation-based (a) and voxel rearrangement-based (b) approaches.}
     \label{fig:mesh}
\end{figure*}

\autoref{fig:mesh} shows a comparison of the triangulation results on the completed skull grids from the interpolation-based (a) and voxel rearrangement-based (b) approaches. It is evident that the mesh from the voxel-rearrangement based method is smooth, while the skull mesh from the interpolation-based method has a bumpy surface.

\begin{figure*}[ht]
     \centering
        \includegraphics[width=\textwidth]{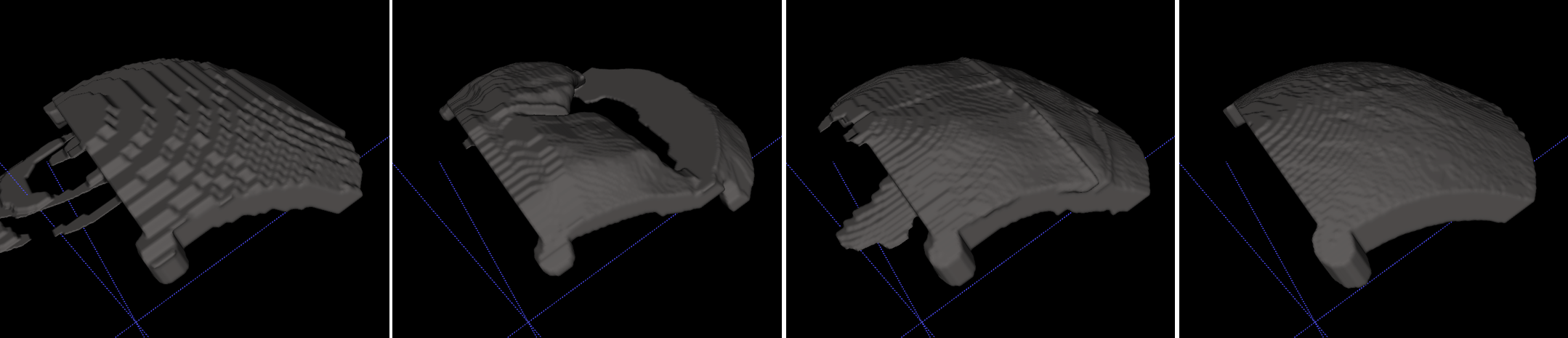}
     \caption{From left to right: implant (voxel grid) obtained from the interpolation-based method, patch-wise skull shape completion and voxel rearrangement-based method. Rightmost: the ground truth implant (a test case selected from $D_{100}$).}
     \label{fig:imp_cmp}
\end{figure*}

\begin{figure*}[ht]
     \centering
        \includegraphics[width=\textwidth]{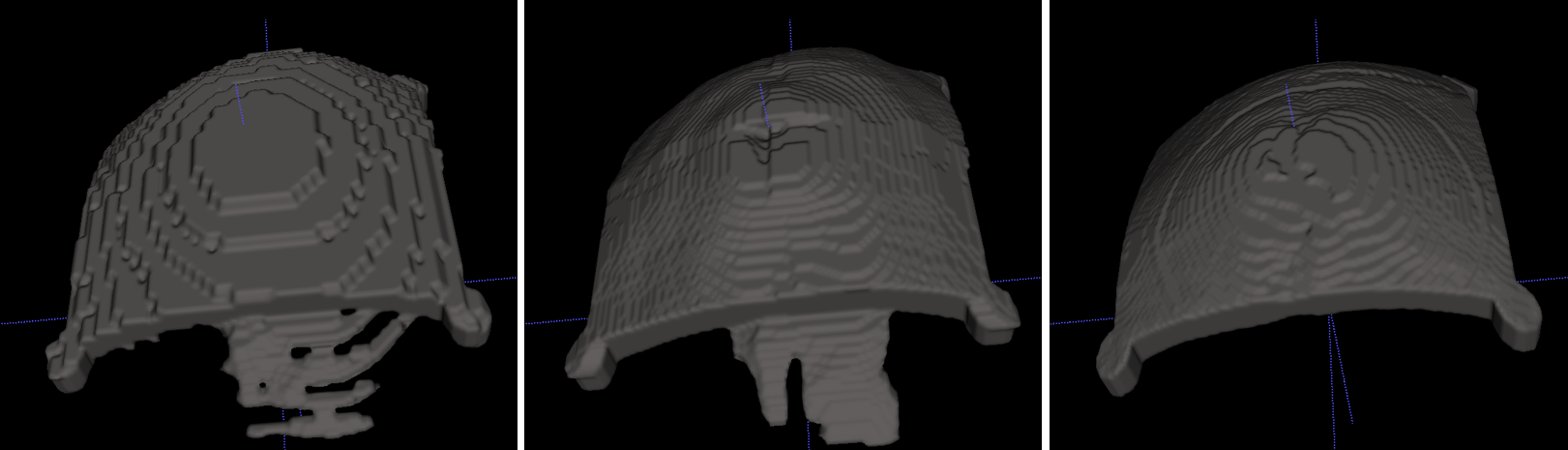}
     \caption{From left to right: implant (voxel grid) obtained from the interpolation-based method, voxel rearrangement-based method and ground truth (a test case selected from $D_{10}$).}
     \label{fig:imp_cmp_special}
\end{figure*}

\autoref{fig:imp_cmp} and \autoref{fig:imp_cmp_special} show the implants produced by the three approaches\footnote{The visualized implants in these figures, as well as the implants used for calculating the DSCs and HDs (Table \ref{table:table1}, \autoref{fig:dsc_hd}) are post-processed using the \textit{denoise} script from the repository: \url{https://github.com/Jianningli/voxel_rearrangement}.}. The patch-wise shape completion method failed on $D_{10}$ and therefore, the result is not shown in \autoref{fig:imp_cmp_special}. We can see that post-processing using connected component analysis and morphological operations did not fully remove the artifacts on the implant. These artifacts (e.g., the non-implant piece on the implant borders in \autoref{fig:imp_cmp} and \autoref{fig:imp_cmp_special}) can substantial increase the Hausdorff Distance (HD) between the prediction and ground truth, as reported in Table \ref{table:table1}. The proposed voxel rearrangement-based method has the best performance on the implants regarding DSC.

\begin{figure*}[ht]
     \centering
        \includegraphics[width=\textwidth]{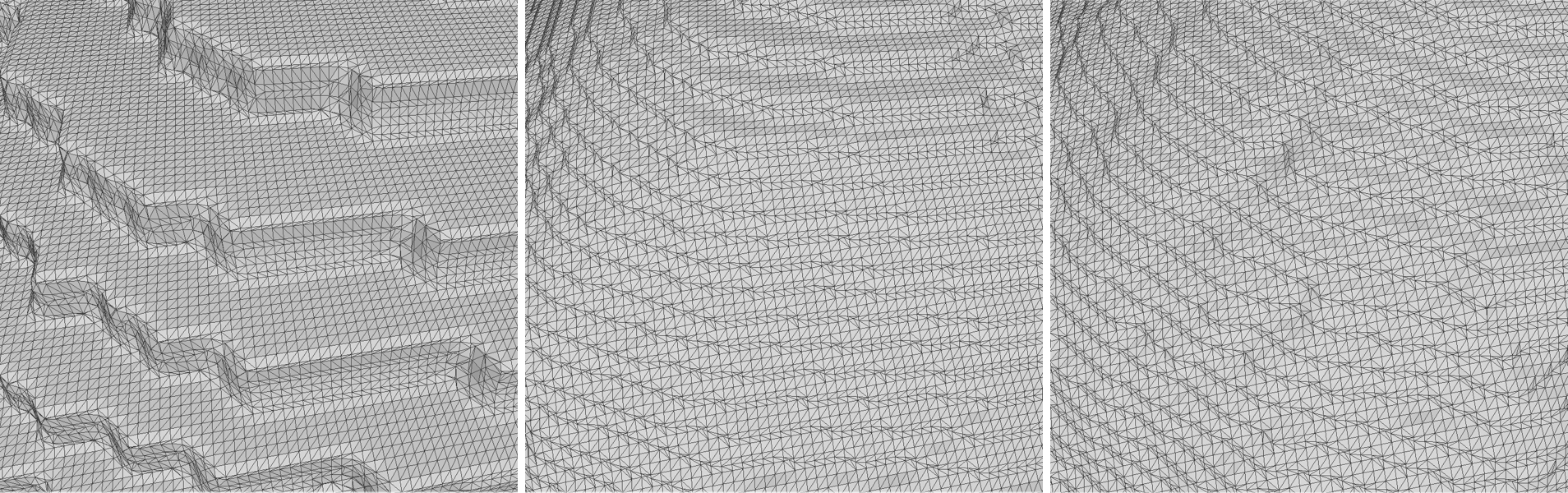}
     \caption{Triangulation results on the implant voxel grid. From left to right: interpolation-based method, voxel rearrangement-based method and ground truth.}
     \label{fig:imp_mesh_cmp}
\end{figure*}

\autoref{fig:imp_mesh_cmp} illustrates the corresponding meshes of the implants shown in \autoref{fig:imp_cmp_special}. We can see that the surface of the mesh from the interpolation-based method has obvious terracing artifacts. 3D printing of such a mesh will yield an implant with a rough surface, which is unusable in cranioplasty. On the contrary, the surface of the mesh from the proposed voxel rearrangement-based method is smooth and close to that of the ground truth.

\subsection{Image Synthesis}

\begin{figure*}
     \centering
        \includegraphics[width=\textwidth]{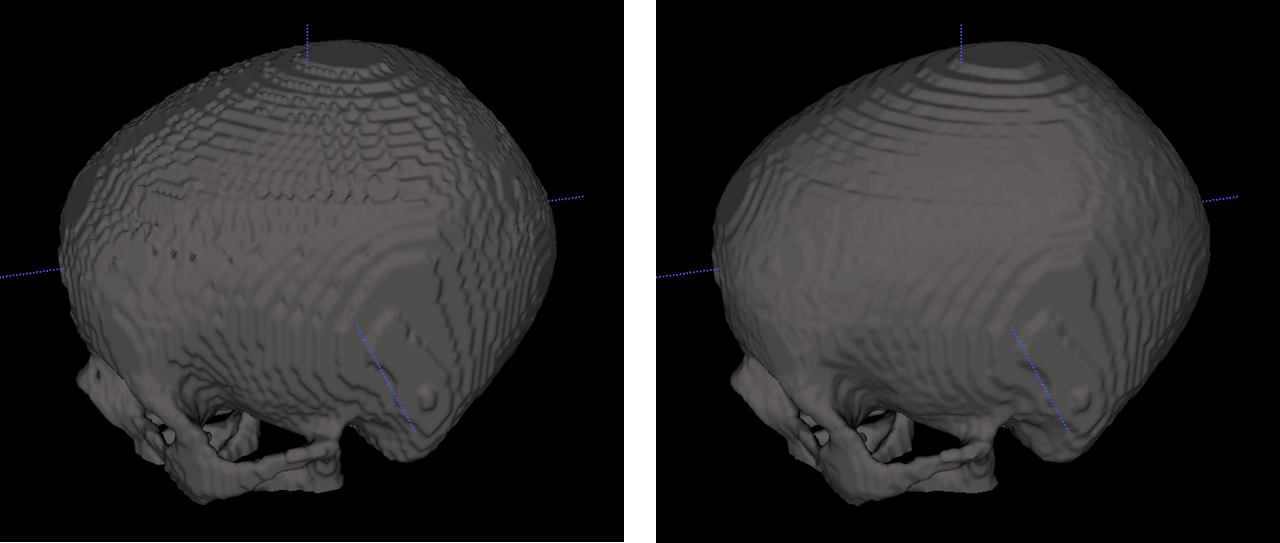}
     \caption{High-resolution complete skull produced by the hash table-based (left) and kd-tree-based (right) image synthesis methods. Note: the skulls are synthesized from size $128 \times 128 \times 64$ to size $256 \times 256 \times 128$.}
     \label{fig:synthesis_result}
\end{figure*}

As the image synthesis-based upsampling only works for images of certain sizes ($2^n$, e.g., $512 \times 512 \times 128$ or $512 \times 512 \times 256$), its quantitative performance (DSC and HD) is not evaluated on the whole test set and, therefore, not comparable to the other methods as reported in Table \ref{table:table1} and \autoref{fig:dsc_hd}. However, qualitative results and comparison with a kd-tree-based image synthesis method for image upsampling will be given in this section.  

Different from the proposed hash table-based image synthesis (\autoref{fig:image_synthesis} (b)), the kd-tree-based method does not employ a binary encoding of the voxels. Instead, principal component analysis (PCA) is used to reduce the dimension of the feature vector (e.g., the $3^3=27$ neighboring voxels) to 20 in order to accelerate the search. The feature vectors from the template pyramid are used to construct a kd-tree and, correspondingly, the feature vectors from the coarse pyramid are also projected into the same principal component space. The NNS can, therefore, be performed on a kd-tree structure, which is by magnitudes faster than a brute-force linear search strategy. It should be noted that the feature vectors will lose binariness after PCA is applied. Unlike the proposed hash table-based method, which performs essentially an approximated NNS, the kd-tree based method performs an exact NNS. 

\autoref{fig:synthesis_result} shows a comparison of the skulls from two image synthesis methods\footnote{The underlying network used to produce the coarse completed skull ($128 \times 128 \times 64$) is the same as that of the interpolation and voxel rearrangement-based method.}. We can see the skull from the kd-tree-based method is of higher quality, which is understandable, as the kd-tree-based method does an exact NNS, whilst the hash table-based method does an approximated NNS. However, the memory consumption of the hash table-based method is much lower \footnote{e.g., 12 \% and 59\% CPU memory consumption for the hash table and kd-tree-based method, respectively.}, as the binary feature vectors are stored as bit strings. Both approaches take less than three minutes to update all the required voxels for each level of the pyramid. Note that the synthesis process in this study is implemented to run on CPUs only. On GPUs, where the number of computing cores is usually large, the process could potentially be several magnitudes faster.

\section{Discussion and Future Work}
3D medical images are known to consume substantial computational resources (e.g., memory, number of FLOPs, etc.) in deep learning. By exploiting the characteristics, e.g., binariness and sparsity, of some specific images, computation requirement for processing these images can be effectively reduced. In our study, we have designed a tailored hash table-based method for nearest neighbor search, by making use of the binariness and spatial sparsity of the high-resolution skull images. However, the template skull image is selected randomly, which is suboptimal considering that the possibility of finding an exact match ($d_{HM}=0$) in the template pyramid might be low for some cases, making the performance of the method unstable. In \cite{dai2017shape}, the authors proposed to select the template image by matching each specific coarse image with a pool of candidate templates in a feature space of a trained autoencoder network. The template that has the smallest distance to the coarse image can be selected and used as the final template for synthesis. For our future work, using the same strategy\footnote{The \textit{feature space} script can be found in the following GitHub repository: \url{https://github.com/Jianningli/voxel_rearrangement}.} could potentially maximize the likelihood of finding the exact bit strings in the template pyramid for each bit string in the coarse template, which consequently increases the final reconstruction accuracy and would probably stabilize the performance of the method. It should be noted that the image synthesis-based method for image upsampling has not been fully evaluated in this study, partly because the method cannot synthesize arbitrarily-sized images. Another reason is that, while Algorithm 1, which lies at the core of the method, is stable, how to create the pyramid (\autoref{fig:image_synthesis} (a)) and the hash table entries still remains open for discussion.  

The voxel rearrangement-based method, on the contrary, has been sufficiently evaluated in our study. One unsolved issue for the method is the \textit{stitching} lines between the contacting borders of patches. For future work, this problem could be solved by using the entire original skull images as the input of the network, instead of using a patch-wise training strategy. Tailored network architectures such as sparse convolutional neural networks could be designed to tackle the memory issue, taking into consideration the binariness and spatial sparsity of the skull data.

\section{Conclusion}
In our study, we have addressed the problem of high memory consumption of 3D medical images. Instead of processing the original high-resolution images directly, high-resolution outcome can be obtained indirectly in a coarse-to-fine fashion. Voxel rearrangement and image synthesis have proven to be effective in restoring the surface smoothness of the coarse output. Both approaches are general and can be used in other applications besides reconstruction, such as medical image segmentation.  \\

\begin{flushleft}
\textbf{Acknowledgement}

This work was supported by the following funding agencies:
\end{flushleft}

\begin{itemize}

\item[$\bullet$] CAMed (COMET K-Project 871132, see also \url{https://www.medunigraz.at/camed/}), which is funded by the Austrian Federal Ministry of Transport, Innovation and Technology (BMVIT) and the Austrian Federal Ministry for Digital and Economic Affairs (BMDW), and the Styrian Business Promotion Agency (SFG);

\item[$\bullet$] The Austrian Science Fund (FWF) KLI 678-B31 (enFaced).

\end{itemize}

\bibliographystyle{splncs04}
\bibliography{references}

\begin{thebibliography}{10}
\providecommand{\url}[1]{\texttt{#1}}
\providecommand{\urlprefix}{URL }
\providecommand{\doi}[1]{https://doi.org/#1}

\bibitem{akil2020fully}
Akil, M., Saouli, R., Kachouri, R., et~al.: Fully automatic brain tumor
  segmentation with deep learning-based selective attention using overlapping
  patches and multi-class weighted cross-entropy. Medical image analysis
  \textbf{63},  101692 (2020)

\bibitem{chang2021three}
Chang, Y.Z., Wu, C.T., Yang, Y.H.: Three-dimensional deep learning to
  automatically generate cranial implant geometry  (2021)

\bibitem{dai2017shape}
Dai, A., Ruizhongtai~Qi, C., Nie{\ss}ner, M.: Shape completion using
  3d-encoder-predictor cnns and shape synthesis. In: Proceedings of the IEEE
  Conference on Computer Vision and Pattern Recognition. pp. 5868--5877 (2017)

\bibitem{diaz2017downsampling}
D{\'\i}az~Garc{\'\i}a, J., Brunet~Crosa, P., Navazo~{\'A}lvaro, I.,
  V{\'a}zquez~Alcocer, P.P.: Downsampling methods for medical datasets. In:
  Proceedings of the International conferences Computer Graphics,
  Visualization, Computer Vision and Image Processing 2017 and Big Data
  Analytics, Data Mining and Computational Intelligence 2017: Lisbon, Portugal,
  July 21-23, 2017. pp. 12--20. IADIS Press (2017)

\bibitem{graham20183d}
Graham, B., Engelcke, M., Van Der~Maaten, L.: 3d semantic segmentation with
  submanifold sparse convolutional networks. In: Proceedings of the IEEE
  conference on computer vision and pattern recognition. pp. 9224--9232 (2018)

\bibitem{graham2017submanifold}
Graham, B., van~der Maaten, L.: Submanifold sparse convolutional networks.
  arXiv preprint arXiv:1706.01307  (2017)

\bibitem{han2017high}
Han, X., Li, Z., Huang, H., Kalogerakis, E., Yu, Y.: High-resolution shape
  completion using deep neural networks for global structure and local geometry
  inference. In: Proceedings of the IEEE international conference on computer
  vision. pp. 85--93 (2017)

\bibitem{kodym2020cranial}
Kodym, O., {\v{S}}pan{\v{e}}l, M., Herout, A.: Cranial defect reconstruction
  using cascaded cnn with alignment. In: Cranial Implant Design Challenge. pp.
  56--64. Springer (2020)

\bibitem{kodym2020skull}
Kodym, O., {\v{S}}pan{\v{e}}l, M., Herout, A.: Skull shape reconstruction using
  cascaded convolutional networks. Computers in Biology and Medicine
  \textbf{123},  103886 (2020)

\bibitem{li2021automatic}
Li, J., von Campe, G., Pepe, A., Gsaxner, C., Wang, E., Chen, X., Zefferer, U.,
  T{\"o}dtling, M., Krall, M., Deutschmann, H., et~al.: Automatic skull defect
  restoration and cranial implant generation for cranioplasty. Medical Image
  Analysis p. 102171 (2021)

\bibitem{li2020dataset}
Li, J., Egger, J.: Dataset descriptor for the autoimplant cranial implant
  design challenge. In: Cranial Implant Design Challenge. pp. 10--15. Springer
  (2020)

\bibitem{li2020baseline}
Li, J., Pepe, A., Gsaxner, C., von Campe, G., Egger, J.: A baseline approach
  for autoimplant: the miccai 2020 cranial implant design challenge. arXiv
  preprint arXiv:2006.12449  (2020)

\bibitem{li2021autoimplant}
Li, J., Pimentel, P., Szengel, A., Ehlke, M., Lamecker, H., Zachow, S.,
  Estacio, L., Doenitz, C., Ramm, H., Shi, H., et~al.: Autoimplant 2020-first
  miccai challenge on automatic cranial implant design. IEEE Transactions on
  Medical Imaging  (2021)

\bibitem{riegler2017octnet}
Riegler, G., Osman~Ulusoy, A., Geiger, A.: Octnet: Learning deep 3d
  representations at high resolutions. In: Proceedings of the IEEE conference
  on computer vision and pattern recognition. pp. 3577--3586 (2017)

\end{thebibliography}
\end{document}